\documentclass[a4paper,11pt]{article}
\pdfoutput=1 
\usepackage{jcappub} 
\usepackage[T1]{fontenc} 
\usepackage{physics}

\usepackage{xcolor}
\usepackage{graphicx}
\usepackage{caption}
\usepackage{subcaption}
\usepackage[export]{adjustbox} 
\usepackage{booktabs}

\usepackage{orcidlink}

\usepackage[nameinlink,noabbrev]{cleveref}
\crefname{equation}{Eq.}{Eqs.}
\crefname{section}{Section}{Sections}
\crefname{figure}{Figure}{Figures}
\crefname{table}{Table}{Tables}
\crefname{appendix}{Appendix}{Appendices}
\Crefname{figure}{Figure}{Figures}
\Crefname{equation}{Equation}{Equations}
\Crefname{section}{Section}{Sections}
\Crefname{table}{Table}{Tables}

\usepackage{lineno}

\title{\boldmath Blinding scheme for the scale-dependence bias signature of local primordial non-Gaussianity for DESI 2024}

\affiliation{Affiliations are in Appendix \ref{sec:affiliations}}

\author[1,2]{{E.~Chaussidon}\orcidlink{0000-0001-8996-4874},}
\author[2]{{A.~de~Mattia}\orcidlink{0000-0003-0920-2947},}
\author[2]{{C.~Yèche}\orcidlink{0000-0001-5146-8533},}
\author[1]{{J.~Aguilar},}
\author[3]{{S.~Ahlen}\orcidlink{0000-0001-6098-7247},}
\author[4]{{D.~Brooks},}
\author[1]{{T.~Claybaugh},}
\author[5]{{S.~Cole}\orcidlink{0000-0002-5954-7903},}
\author[6]{{A.~de la Macorra}\orcidlink{0000-0002-1769-1640},}
\author[4]{{P.~Doel},}
\author[7,8]{{K.~Fanning}\orcidlink{0000-0003-2371-3356},}
\author[9,10,11]{{E.~Gaztañaga},}
\author[1]{{S.~Gontcho A Gontcho}\orcidlink{0000-0003-3142-233X},}
\author[12]{{C.~Howlett}\orcidlink{0000-0002-1081-9410},}
\author[1]{{T.~Kisner}\orcidlink{0000-0003-3510-7134},}
\author[1]{{A.~Lambert},}
\author[13]{{L.~Le~Guillou}\orcidlink{0000-0001-7178-8868},}
\author[14,15]{{M.~Manera}\orcidlink{0000-0003-4962-8934},}
\author[16]{{A.~Meisner}\orcidlink{0000-0002-1125-7384},}
\author[17,15]{{R.~Miquel},}
\author[18,19]{{G.~Niz}\orcidlink{0000-0002-1544-8946},}
\author[2,1]{{N.~Palanque-Delabrouille}\orcidlink{0000-0003-3188-784X},}
\author[20,21,22]{{W.~J.~Percival}\orcidlink{0000-0002-0644-5727},}
\author[23]{{F.~Prada}\orcidlink{0000-0001-7145-8674},}
\author[24,25,26]{{A.~J.~Ross}\orcidlink{0000-0002-7522-9083},}
\author[27]{{G.~Rossi},}
\author[28]{{E.~Sanchez}\orcidlink{0000-0002-9646-8198},}
\author[1]{{D.~Schlegel},}
\author[29,30]{{M.~Schubnell},}
\author[31]{{H.~Seo}\orcidlink{0000-0002-6588-3508},}
\author[16]{{D.~Sprayberry},}
\author[30]{{G.~Tarl\'{e}}\orcidlink{0000-0003-1704-0781},}
\author[6]{{M.~Vargas-Maga\~na}\orcidlink{0000-0003-3841-1836},}
\author[16]{{B.~A.~Weaver},}
\author[32]{{H.~Zou}\orcidlink{0000-0002-6684-3997},}

\emailAdd{echaussidon@lbl.gov}

\abstract{The next generation of spectroscopic surveys is expected to achieve an unprecedented level of accuracy in the measurement of cosmological parameters. To avoid confirmation bias and thereby improve the reliability of these results, blinding procedures become a standard practice in the cosmological analyses of such surveys. Blinding is especially crucial when the impact of observational systematics is important relative to the cosmological signal, and a detection of that signal would have significant implications. This is the case for local primordial non-gaussianity, as probed by the scale-dependent bias of the galaxy power spectrum at large scales that are heavily sensitive to the dependence of the target selection on the imaging quality, known as imaging systematics. We propose a blinding method for the scale-dependent bias signature of local primordial non-gaussianity at the density field level which consists in generating a set of weights for the data that replicate the scale-dependent bias. The applied blinding is predictable, and can be straightforwardly combined with other catalog-level blinding procedures that have been designed for the baryon acoustic oscillation and redshift space distortion signals. The procedure is validated through simulations that replicate data from the first year of observation of the Dark Energy Spectroscopic Instrument, but may find applications to other upcoming spectroscopic surveys.}

\begin{document}
\maketitle
\flushbottom

\section{Introduction}
In addition to providing tight constraints on the Universe expansion history, one of the major roles of current \cite{DESICollaboration2016, Scaramella2022} or future spectroscopic surveys \cite{Dore2014, Ferraro2019} is to provide an accurate measurement of the amount of primordial non-gaussianity (PNG) generated during inflation. In particular, these surveys will be focused on the well-known local PNG parameterized by $f_{\mathrm{NL}}^{\mathrm{loc}}$ \cite{Komatsu2001} such that the primordial gravitational potential field $\Phi(\vb{x})$ reads
\begin{equation}
 \Phi(\vb{x}) =  \Phi_g(\vb{x}) + f_{\mathrm{NL}}^{\mathrm{loc}} \left( \Phi_g^2(\vb{x}) - \langle \Phi_g^2 \rangle_{\vb{x}} \right),
\end{equation}
where $\Phi_g$ is the Gaussian primordial potential field. A detection of $\vert f_{\mathrm{NL}}^{\mathrm{loc}} \vert \gg 0.01$ would immediately rule out single field slow-roll inflation \cite{Creminelli2004} as a valid paradigm to describe the early Universe. 

Currently, the best constraints on local PNG are obtained from Planck data: $f_{\mathrm{NL}}^{\mathrm{loc}} = -0.9 \pm 5.1~(68\%)$ \cite{Akrami2020}. To circumvent the cosmic variance limit of CMB observations, spectroscopic surveys use the enormous statistical power in the 3D galaxy clustering, probing a large volume of the Universe. A promising approach to constrain local PNG is through detecting the tiny imprint it leaves on the galaxy power spectrum at large scales, known as the scale-dependent bias on large scales \cite{Dalal2008, Slosar2008}. With this method, the best measurement leads to $ -23 < f_{\mathrm{NL}}^{\mathrm{loc}} < 21~(68\%)$ \cite{Castorina2019, Mueller2021, Cagliari2023} and is obtained with the quasars of the extended Baryon Oscillation Spectroscopic Survey (eBOSS) \cite{Myers2015}. 

Current spectroscopic surveys, as the Dark Energy Spectroscopic Instrument (DESI) \cite{DESICollaboration2016,Abareshi2022} or Euclid \cite{Scaramella2022}, are expected to constrain local PNG with similar accuracy to Planck. At the same time, future surveys are expected to reach a sensitivity of order unity \cite{Heinrich2023}, requiring robust and confirmation bias-free analyses.

Besides, the large-scale modes of the power spectrum used in the scale-dependent bias measurement may be strongly impacted by angular modes resulting from the dependence of the target selection to the imaging quality \cite{Rezaie2021}. This dependence, known as \emph{imaging systematics}, was the object of many studies carried out for the Sloan Digital Sky Survey (SDSS) \cite{Myers2006, Ross2011, ROss2017, Ross2020, Ho2012,Kalus2019,Raichoor2021,Kong2020}, the Dark Energy Survey (DES) \cite{Leistedt2016,Suchyta2016,Elvin-Poole2018,Weaverdyck2021,Everett2022}, or more recently to prepare the upcoming DESI clustering analysis \cite{Kitanidis2020,Rezaie2019,Chaussidon2022}. Correction schemes mostly rely on template fitting methods which need to be carefully calibrated to mitigate imaging systematics while avoiding overfitting. Overfitting indeed leads to a deficit of power at large scales that biases the analysis \cite{Krolewski2023,Rezaie2023}. Unfortunately, the efficiency of the systematic mitigation is typically validated on the actual data through several summary statistics, including the power spectrum itself, such that one may be influenced by previous detection (or non-detection) of PNG when tuning the systematic corrections, resulting in an important confirmation bias. 

Similarly to the seminal blinding method for the baryon acoustic oscillation (BAO) and redshift-space distortion (RSD) analyses \cite{Brieden2020}, we propose here a method that introduces at the catalog level a fake signal mimicking the scale-dependent bias that allows us to investigate the large scales of the power spectrum while protecting ourselves from confirmation bias. \cref{sec:method} describes the blinding scheme. Then, in \cref{sec:test}, we validate the impact of this method on the power spectrum. In \cref{sec:realistic_test}, we perform realistic tests with non-zero PNG simulations and validate the combination of the blinding method with the imaging systematic mitigation. We conclude in \cref{sec:conclusion}.

\section{Theoretical description} \label{sec:method}
In the following section we first introduce the signature of local PNG that we aim at blinding. We then describe the blinding scheme and its practical implementation, noting a shot noise correction is required to remove the sensitivity of the blinding scheme with the the tracer density.

\subsection{Scale-dependent bias}
Local primordial non-gaussianity leaves an imprint on the tracer power spectrum at large scales known as the scale-dependent bias \citep{Dalal2008, Slosar2008}:
\begin{equation} \label{eqn:scale_depedent_bias}
    P(k, z) = \left(b(z) + \dfrac{b_{\Phi}(z)}{\alpha(k, z)} f_{\mathrm{NL}}^{\mathrm{loc}} \right)^2 P_{\mathrm{lin}}(k, z),
\end{equation}
where $P_{\mathrm{lin}}$ is the linear matter power spectrum, $\alpha(k, z)$ is a transfer function connecting the primordial gravitational field $\Phi$ to the matter density perturbation such that 
\begin{equation}
    \delta(k, z) = \alpha(k, z) \Phi(k),
\end{equation}
$b(z)$ is the linear bias of the tracer and $b_{\Phi}$ is the PNG bias given the response to the presence of local PNG of the tracer. 

The transfer function $\alpha(k, z)$ can be computed directly as\footnote{We use \texttt{CLASS} \cite{Blas2011} wrapped by \url{https://github.com/cosmodesi/cosmoprimo}}:
\begin{equation}
\alpha(k, z) = T_{\Phi \rightarrow \delta}(k, z) = \sqrt{\dfrac{P_{\delta}(k, z)}{P_{\Phi(k)}}},
\end{equation}
where the primordial potential\footnote{$\Phi$ is normalized to $3/5 \mathcal{R}$ to match the usual definition of \cite{Slosar2008}.} 
power spectrum $P_{\Phi}$ is defined as:
\begin{equation}
P_{\Phi}(k) = \dfrac{9}{25} \dfrac{2 \pi^2}{k^3} A_s \left(\dfrac{k}{k_{\rm pivot}}\right)^{n_s-1},
\end{equation}
with $n_s$ is the spectral index and $A_s$ the amplitude of the initial power spectrum at $k_{\rm pivot} = 0.05\;\text{Mpc}$. $P_{\delta}$ is the linear matter power spectrum such that $\alpha(k, z)$ has the famous scale dependency: $ \alpha(k, z) \propto  k^2 \times T_{\Phi \rightarrow \Phi}(k, z)$ \cite{Dalal2008}.

The PNG bias $b_{\Phi}$ can be expressed as $b_{\Phi}(z) = 2 \delta_c \times (b(z) - p)$ with $p=1$~\cite{Slosar2008} (universal mass function) for tracers whose halo occupation distribution depends only on mass, although the pertinence of this description is currently discussed \cite{Barreira2022, Fondi2023, Sullivan2023}. Since $b_{\Phi}$ and $f_{\mathrm{NL}}^{\mathrm{loc}}$ are fully degenerate, in the following we assume the universal mass function to fix $b_{\Phi}$ and we express the blinding amplitude in terms of $f_{\mathrm{NL}}^{\mathrm{loc}}$.

\subsection{Blinding scheme} 
\subsubsection{Mimicking the scale-dependent bias} \label{sec:model}

Our proposed blinding consists in adding a fake signal to the galaxy catalog, such that the measured power spectrum matches the large scale-dependent bias given in \cref{eqn:scale_depedent_bias} for a specific value of $f_{\mathrm{NL}}^{\mathrm{loc}}$, which we dub $f_{\mathrm{NL}}^{\mathrm{blind}}$. This value is fixed randomly and is to remain unknown for all the data consistency tests performed during the analysis. Hence, the expected amount of PNG on the blinded data (assuming the chosen $b_{\Phi}$ for the blinding is the true one) is $f_{\mathrm{NL}}^{\mathrm{loc}} + f_{\mathrm{NL}}^{\mathrm{blind}}$ where $f_{\mathrm{NL}}^{\mathrm{loc}}$ is the true value of the underlying cosmology and $f_{\mathrm{NL}}^{\mathrm{blind}}$ a random, fixed and unknown value.

Analyses of current spectroscopic surveys traditionally weight the data to correct for several observational effects as the completeness of the observations, the imaging dependence of the target selection or even the spectroscopic efficiency, see \cite{Ross2024} for the description of the weights used in DESI. 
We alter the measured power spectrum at large scales by including in the catalog an additional set of weights, multiplied by the traditional ones to make them indistinguishable, such that $\delta_g \rightarrow \delta_g + a(k) \delta^r$, where $\delta_g$ is the galaxy density field and $\delta^r$ is the real space density field. The proposed weights are
\begin{equation}
\label{eqn:weights_data}
    w_{\phantom{d}\mathrm{blind}}^{d}(\vb{x}) = 1 + w_{\mathrm{blind}}(\vb{x})
\end{equation}
where $w_{\mathrm{blind}}(\vb{x})$ is the Fourier transform of
\begin{equation}  \label{eqn:weights_correction}
    w_{\mathrm{blind}}(\vb{k}) =  \dfrac{b_{\Phi}f_{\mathrm{NL}}^{\mathrm{blind}}}{b \alpha(k)} \times \widehat{\delta}^r(\vb{k}) \equiv a(k) \times \widehat{\delta}^r(\vb{k}).
\end{equation}
and $\widehat{\delta}^r$ is an estimate of $\delta^r$. 
The quantities $b$, $b_{\Phi}$ and $\alpha(k)$ are computed at the effective redshift $z_{\mathrm{eff}}$ of the data (see, for instance, Eq. (3.5) of \cite{Cagliari2023}). A wrong effective redshift will increase or decrease the real amount of added fake signal, such that the apparent value of $f_{\mathrm{NL}}^{\mathrm{blind}}$ will be slightly different. Note that a wrong bias will also alter the estimation of the real space density field. However all of these impacts will be neglected in the following, since we will use the correct effective redshift. Let us now detail the implementation of this blinding scheme.

\subsubsection{Computing blinding weights}
\label{sec:computing_blinding_weights}

The real space density field $\widehat{\delta}^r(\vb{k})$ in~\cref{eqn:weights_correction} is estimated by performing a reconstruction process, see~\cite{Paillas2024} for a complete description. First, the density field is built by painting the particles (data and randoms) on a grid using the cloud-in-cell (CIC) assignment scheme. The smoothed Fourier-space density field $\widehat{\delta}(\mathbf{k})$ is obtained with a Fast Fourier Transform (FFT) and a product with the Gaussian kernel
\begin{equation}
S_{1}(\vb{k}) = \exp(-\frac{1}{2}k^2\sigma_{1}^2).
\end{equation}

RSD displacements are then estimated from $\widehat{\delta}(\mathbf{k})$, assuming a fiducial bias $b$ and growth rate $f$ with the iterative FFT reconstruction algorithm \citep{Burden2018} implemented in \texttt{pyrecon}\footnote{\url{https://github.com/cosmodesi/pyrecon/tree/mpi}. This branch is an MPI implementation and this is the only difference with the main branch.}. Data particles are shifted back by the opposite of the RSD displacements, painted on a new grid, which is again Fourier-transformed, divided by $b$ and multiplied by another smoothing kernel $S_{2}(\vb{k}) = \exp(-\frac{1}{2}k^2\sigma_{2}^2)$ to obtain an estimate of $\widehat{\delta}^r(\vb{k})$. 

Finally, $w_{\phantom{d}\mathrm{blind}}^{d}(\vb{x})$ is read at the galaxy positions with the CIC scheme from the FFT of $a(k) \widehat{\delta}^r(\vb{k})$. 

\begin{figure}
    \centering
    \includegraphics[scale=1]{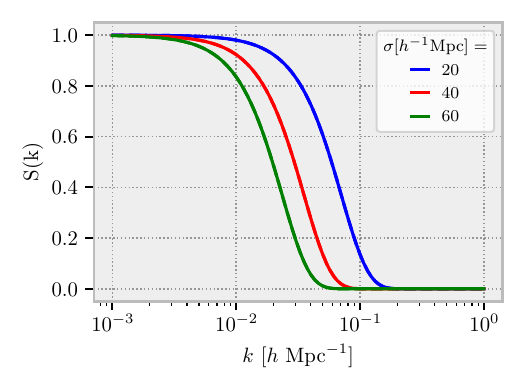}
    \caption{Smoothing kernel for several values of smoothing radius parameter $\sigma$.}
    \label{fig:window_reconstrucion}
\end{figure}

Since we only need to reconstruct the real-space density field at large scales (where PNG matters), we use a large smoothing radius $\sigma_{1} = \sigma_{2} = 30~h^{-1}\textrm{Mpc}$. The corresponding kernel is displayed in \cref{fig:window_reconstrucion} for several values of $\sigma$. The larger the smoothing parameter, the more the fluctuations of the reconstructed density field on small scales (large $k$) is suppressed. The smoothing parameter $\sigma_2$ allows us to control which scales will be affected by this blinding scheme, thereby preventing the BAO or RSD scales from being contaminated by these additional weights. While, the smoothing parameter $\sigma_1$ controls the scale where the reconstruction is effective, see \cref{eqn:y_tilde}. More details on the effect of smoothing will be given in \cref{sec:lower_bias}.

\subsubsection{From galaxy catalog to the FKP field}
Let us now describe the impact of the blinding weights obtained above on the estimated galaxy power spectrum.
The FKP field \cite{Feldman1994}, including the blinding weights above, reads:
\begin{align}  \label{eqn:FKP_field}
\begin{split}
    F(\vb{x}) &= n_g(\vb{x}) - \alpha_s n_s(\vb{x}) \\
         &= W(\vb{x}) \left(1 + \delta_g(\vb{x})\right) \left( 1 + w_{\mathrm{blind}}(\vb{x}) \right) - W(\vb{x}) \\ 
         &\simeq W(\vb{x}) \delta_g(\vb{x}) + W(\vb{x}) w_{\mathrm{blind}}(\vb{x}), 
\end{split}
\end{align}
where we used the usual notation: $\alpha_s = (\int \dd \vb{x} W(\vb{x})) /(\int \dd \vb{x} n_s(\vb{x}))$ and $n_g$ (\textit{resp.} $n_s$) is the galaxy (\textit{resp.} randoms) density. The randoms samples the survey geometry, more specifically the \emph{survey selection function} $W(\vb{x})$, which is the ensemble average of the galaxy density: $W(\vb{x})=\left\langle n_g(\vb{x})\right\rangle = \langle \alpha_s n_s(\vb{x}) \rangle$.

In \cref{eqn:FKP_field}, the term $\delta_g(x) w_{\mathrm{blind}}(x)$ can be neglected since it is a second-order term in $\delta_g$. As described in \cref{sec:weights_data_randoms}, this assumption can be avoided by weighting the randoms instead of the data, though our tests have shown no practical difference.

One can also remark that, under this approximation, the expected value of $n_g$ is unchanged with the blinding scheme applied to the data. Indeed, the expected value reads:
\begin{align}
\begin{split}
      \langle n_{g}(\vb{x})\rangle &= W(\vb{x}) \langle 1 + \delta_{g}(\vb{x}) + w_{\mathrm{blind}}(\vb{x}) + \delta_{g}(\vb{x}) w_{\mathrm{blind}}(\vb{x})\rangle \\
      &\simeq W(\vb{x}) \langle 1 + \delta_{g}(\vb{x}) + w_{\mathrm{blind}}(\vb{x}) \rangle \\
      &= W(\vb{x}),
\end{split}
\end{align}
where $<w_{\mathrm{blind}}> = 0$ as the result of the application of the linear operator $a(k)$ to $\delta^r(k)$, of zero ensemble average. 

This provides a simple way to test this approximation by computing the change in $\langle n_{g}(\vb{x}) \rangle$, which corresponds to the change in mean data weights. This change is minimal, typically about $10^{-4}$; see for instance the data blinding weights displayed in \cref{fig:weights_dispersion}.

\subsection{Shot noise correction} \label{sec:shotnoise_correction}
\subsubsection{Power spectrum with the blinding scheme}
Applying the blinding weights~\cref{eqn:weights_data}, the Fourier-space density field becomes ${\delta^{s}}^{'}(\vb{k}) = \delta^s(\vb{k}) + a(k) \widehat{\delta}^r(\vb{k})$ where $\delta^{s}$ is the density field in redshift space and $\widehat{\delta}^{r}$ is the estimation of the field in real space. The associated power spectrum is, in the linear regime:
\begin{equation}\label{eqn:power_spectrum_with_blinding_weights}
    \langle \vert {\delta^{s}}^{'}(\vb{k}) \vert^2 \rangle = \underbrace{\langle \vert \delta^{s}(\vb{k}) \vert^2 \rangle}_{\begin{matrix}\rightarrow (b + f \mu^2)^2 P_{\mathrm{lin}}(k) + 1 / n\end{matrix}} + \quad 2 a(k) Re(\underbrace{\langle \delta^s(\vb{k})\widehat{\delta}^{r\star}(\vb{k})\rangle}_{\begin{matrix}\rightarrow\tilde{X}(\vb{k}, n)\end{matrix}})\quad  + \quad  a(k)^2 \underbrace{\langle \vert \widehat{\delta}^r(\vb{k})\vert^2 \rangle}_{\begin{matrix}\rightarrow\tilde{Y}(\vb{k}, n)\end{matrix}}.
\end{equation}
where $1/n$ is the shot noise. 

Noting that, as described in~\cref{sec:computing_blinding_weights}, $\widehat{\delta}^{r}(\vb{k})$ is an estimate of the reconstructed real space density field using a smoothing kernel $S_{1}(\vb{k})$, and then smoothed by $S_{2}(\vb{k})$,
we find that $\tilde{X}(\vb{k}, n)$ and $\tilde{Y}(\vb{k}, n)$, in the linear regime, are
\begin{align}
    \tilde{X}(\vb{k}, n) &= \left(b + f \mu^2\right) \left(b + \left(1 - S_{1}(k)\right) f \mu^2\right) S_{2}(k) P_{\mathrm{lin}}(k) + \dfrac{1}{n} S_{2}(k) \exp(-\frac{1}{2} k^2 \mu^2 f^2 \sigma_d^2), \label{eqn:x_tilde} \\
    \tilde{Y}(\vb{k}, n) &= \left(b + \left(1 - S_{1}(k)\right) f \mu^2\right)^2 S_{2}(k)^2 P_{\mathrm{lin}}(k) + \dfrac{1}{n} S_{2}(k)^2, \label{eqn:y_tilde}
\end{align}


The shot noise exponential term in \cref{eqn:x_tilde} comes from the displacement between the redshift space and estimated real space density fields, see Eq.(23) of \citep{Smith2021},
with variance $\sigma_d^2$ given by: 
\begin{equation} 
    \sigma_d^2 = \frac{1}{6\pi} \int_0^\infty \dd k S_{1}(k)^{2} P_{\mathrm{lin}}(k).
\end{equation}

At $k > 1 / \sigma_{2}$, the $\tilde{X}(\vb{k}, n)$ and $\tilde{Y}(\vb{k}, n)$ contributions are suppressed by the damping factor $S_{2}(k)$.
At lower $k$, the limit of infinite density ($1/n \rightarrow 0$), \cref{eqn:power_spectrum_with_blinding_weights} converges to the redshift space equivalent of \cref{eqn:scale_depedent_bias}.


\subsubsection{Shot noise contribution} \label{sec:shot_noise_contribution}
The shot noise terms (1/$n$) in \cref{eqn:x_tilde} and in \cref{eqn:y_tilde} yield an extra contribution to the estimated power spectrum. In addition, shot noise $1/n$ varies with the local average density $n$, \textit{i.e.} the survey selection function. In particular, $n$, and hence the shot noise, typically varies with the redshift and across the footprint due to the imaging dependence of the target selection\footnote{See, for instance, \cref{fig:survey_selection_function} for a survey selection function.}. The varying shot noise may then lead to different apparent $f_{\mathrm{NL}}^{\mathrm{loc}}$ values as a function of the redshift bin considered for the power spectrum estimation, preventing us from dedicated study, including internal consistency check on the blinded catalog, such as splitting the sample into several redshift bins. It is therefore crucial to correct for this shot noise effect.

To effectively treat this contribution, let us find $a^{\prime}(\vb{k})$ to recover the expected value of the additional PNG power without the undesired shot noise term:
\begin{equation} \label{eqn:equation_to_solve}
    2 a^{\prime}(\vb{k}) \tilde{X}(\vb{k}, n)  + {a^{\prime}}^2(\vb{k}) \tilde{Y}(\vb{k}, n) = 2 a(k) X(\vb{k}) + a^2(k) Y(\vb{k}) \equiv
 A(\vb{k}),
\end{equation}
where $X(\vb{k})$ (\textit{resp.} $Y(\vb{k})$) corresponds to the zero shot noise part ($1/n \rightarrow 0$) of \cref{eqn:x_tilde} (\textit{resp.} \cref{eqn:y_tilde}). Solving the second-degree polynomial of \cref{eqn:equation_to_solve} gives: 
\begin{equation} \label{eqn:solution}
a^{\prime}(\vb{k}, n) = \dfrac{- \tilde{X}(\vb{k}, n) \pm \sqrt{\tilde{X}(\vb{k}, n)^2 + \tilde{Y}(\vb{k}, n)A(\vb{k})} }{\tilde{Y}(\vb{k}, n)},
\end{equation}
where the choice of the positive or negative solution of \cref{eqn:solution} will be discussed in \cref{sec:positive_negative_solution}.

The new parameter $a^\prime$ depends on both $\vb{k}$ and $\vb{x}$ via the shot noise term in $\tilde{X}(\vb{k}, n)$ and $\tilde{Y}(\vb{k}, n)$. To ensure that the amplitude of the PNG signal generated by the blinding scheme in the power spectrum does not depend on the local survey selection function, we want to keep the dependence on $\vb{x}$ in $a^\prime$ and then for simplicity remove the dependence on $\vb{k}$. Hence, we choose a value of $\vb{k}_p = (k_{p}, \mu_{p})$ where the equality given in \cref{eqn:equation_to_solve} holds and propose new, corrected, weights:
\begin{equation} 
\label{eqn:final_weights}
w_{\phantom{d}\mathrm{blind}}^{d}(\vb{x}) = 1 + \dfrac{a^\prime(k_{p}, \mu_{p}, n(\vb{x}))}{a(k_{p})} w_{\mathrm{blind}}(\vb{x})
\end{equation}
still with $w_{\mathrm{blind}}(\vb{x})$ the Fourier transform of~\cref{eqn:weights_correction} and $a^\prime(k_{p}, \mu_{p}, n) / a(k_{p})$ the corrective factor to suppress the shot noise contribution. 

\subsubsection{Positive or negative solution?} \label{sec:positive_negative_solution}
\begin{figure}  
     \begin{subfigure}{\textwidth}
         \centering
         \includegraphics[scale=1, center]{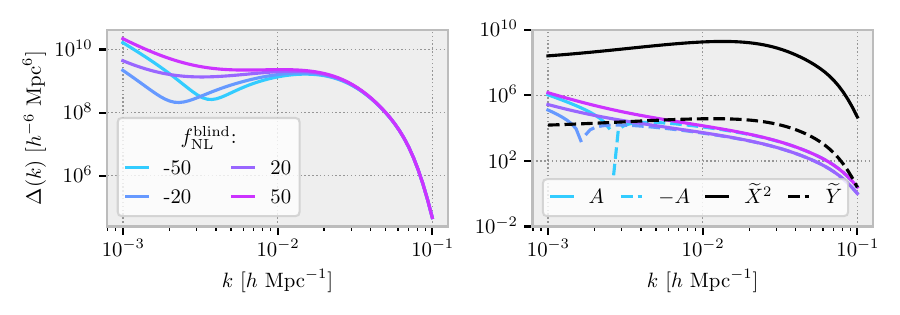}
     \end{subfigure}
     \begin{subfigure}{\textwidth}
         \centering
         \includegraphics[scale=1, center]{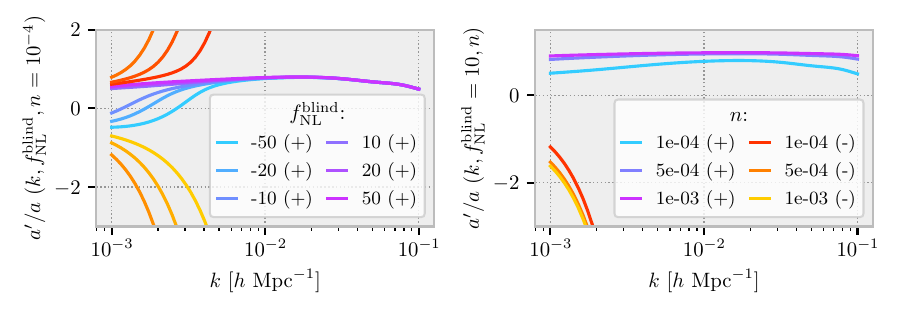}
     \end{subfigure}
     \caption{Numerical study of \cref{eqn:equation_to_solve} and of \cref{eqn:solution}. \textit{Top left:} Reduced discriminant $\Delta(\vb{k}) = \tilde{X}(\vb{k}, n)^2 + \tilde{Y}(\vb{k}, n)A(\vb{k})$ for several values of $f_{\mathrm{NL}}^{\mathrm{blind}}$, $\mu = 0.6$, and shot noise $1/n = 10^{4} \; h^{-3}\textrm{Mpc}^{3}$. The reduced discriminant is always positive, leading to two real solutions. \textit{Top right:} Contribution of the different terms to the reduced discriminant. \textit{Bottom left:} Corrective factor for different values of $f_{\mathrm{NL}}^{\mathrm{blind}}$ with $n = 10^{-4} \; \textrm{Mpc}^{-3}~h^{3}$ and $\mu=0.6$. The $+$ solutions are in blue, and the $-$ ones are in orange. For $f_{\mathrm{NL}}^{\mathrm{blind}} < 0$, we choose the $+$ solution since we fix the value of $k_{\mathrm{p}}$ to $8 \cdot 10^{-3} \; h~\textrm{Mpc}^{-1}$. \textit{Bottom right:} Similar to \textit{Bottom left} but for different values of shot noise with $f_{\mathrm{NL}}^{\mathrm{blind}} = 10$.}
     \label{fig:discriminant}
\end{figure}

The reduced discriminant $\Delta(\vb{k}) = \tilde{X}(\vb{k}, n)^2 + \tilde{Y}(\vb{k}, n)A(\vb{k})$ of \cref{eqn:equation_to_solve} is always positive, as displayed in \cref{fig:discriminant}, such that \cref{eqn:equation_to_solve} has two real solutions.

When $f_{\mathrm{NL}}^{\mathrm{blind}} > 0$, it is clear with \cref{eqn:weights_correction} that we need to choose the positive solution in \cref{eqn:solution} to have $a^\prime(k)$ positive and mimic the positive behavior of local PNG. The choice of the solution in \cref{eqn:solution} for $f_{\mathrm{NL}}^{\mathrm{blind}} < 0$, is motivated to have a corrective factor ($a^{\prime} / a$) that is in $[0, 1]$, since $\tilde{Y}$ is always positive by definition. We want $a^{\prime} / a > 0$ to do not change the sign of the blinding signal and $a^{\prime} / a < 1$ to reduce the amount of the blinding signal in presence of shot noise, see \cref{eqn:power_spectrum_with_blinding_weights}. This corrective factor is plotted in \cref{fig:discriminant} for different values of $f_{\mathrm{NL}}^{\mathrm{blind}}$ and for different values of shot noises. One can check that the corrective factors for the positive solution (blue/violet lines) are in $[0, 1]$ for a reasonable value of $k$ as a function of the value of $f_{\mathrm{NL}}^{\mathrm{blind}}$, while the corrective factors for negative solution (orange lines) do not respect this requirement. 

\begin{figure}
    \centering
    \includegraphics[width=0.9\textwidth, center]{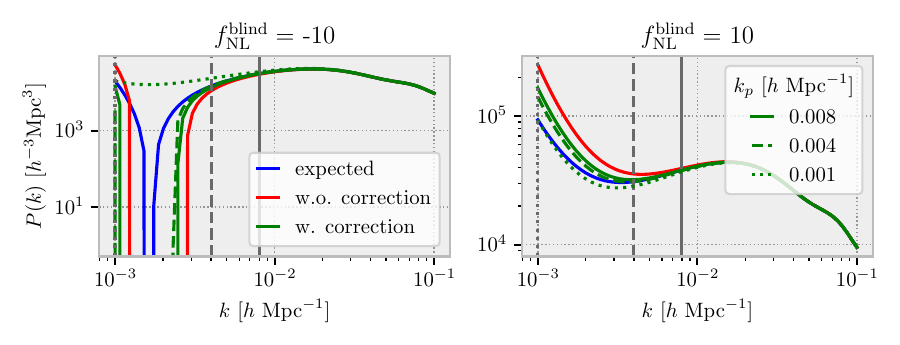}
    \caption{Impact of the value of $k_p$ (with $\mu_p=0.6$) for $f_{\mathrm{NL}}^{\mathrm{blind}}=-10$ (\textit{left}) and for $f_{\mathrm{NL}}^{\mathrm{blind}}=10$ (\textit{right}) on the theoretical description of the power spectrum monopole for a shot noise value of $n = 10^{-4} \; \textrm{Mpc}^{-3}~h^{3}$. The blue lines are the expected power spectrum monopoles, \textit{i.e.}, in the limit $n \rightarrow \infty$. The red lines are given by \cref{eqn:power_spectrum_with_blinding_weights} without the shot noise correction, while the green lines have the correction. \label{fig:impact_correction}}
\end{figure}

As explained in \cref{sec:shot_noise_contribution}, we fix the corrective factor $(a^\prime/a)(\vb{k}, n)$ at a chosen $\vb{k}_p$ value to be in $[0,1]$ for the considered range of blinding. This choice follows from the numerical study in \cref{eqn:solution} and the fact that in the following, we will limit our study to $f_{\mathrm{NL}}^{\mathrm{blind}} \in [-50, 50]$. Note that we expect to use a smaller range during the random choice of the blinding value since, for the upcoming DESI study with the first data release, a statistical precision of $\sigma(f_{\mathrm{NL}}^{\mathrm{loc}})\sim 15$ is expected, such that a lower range of blinding value will be used. Hence, when $f_{\mathrm{NL}}^{\mathrm{blind}}>0$, we choose $k_p = 4 \cdot 10^{-3} \; h~\textrm{Mpc}^{-1}$ in order to de-bias the PNG signal at scales of interest. To avoid potential zero-crossing of the power spectrum when $f_{\mathrm{NL}}^{\mathrm{blind}}$ is negative, we choose a slightly higher value of $k_p = 8 \cdot 10^{-3} \; h~\textrm{Mpc}^{-1}$. We use $\mu_p = 0.6$ in all cases. The impact of the choice of $k_p$ on the theoretical description of the power spectrum given by \cref{eqn:power_spectrum_with_blinding_weights} is shown in \cref{fig:impact_correction}. Note that the $k$-range on which the scale-dependent bias will be fitted with the first data release of DESI will not include modes below $k_{\rm min} \sim 3 \cdot 10^{-3} h~\textrm{Mpc}^{-1}$.

\subsection{Shot noise with data weights}
For the shot noise correction to work, we need an appropriate estimate of the shot noise. We suggest using the same estimation as used in the power spectrum estimation\footnote{\url{https://pypower.readthedocs.io/en/latest/api/api.html\#pypower.fft_power.normalization}}, that is, in each cell of volume $dV$ centered around the position $\vb{x}$:
\begin{equation}
n(\vb{x})^{-1} = \frac{\sum_{dV} w_{g}^{2}}{\alpha_s (\sum_{dV} w_{g}) (\sum_{dV} w_{s}) / dV},
\end{equation}
where $w_g$ (\textit{resp.} $w_s$) represents the weights of the data (\textit{resp.} randoms).

The denominator is an estimation of the integral $\int \dd \vb{x} W^2(\vb{x})$. We use $\alpha_s (\sum_{dV} w_{g}) (\sum_{dV} w_{s})$ instead of $(\sum_{dV} w_{g})^{2}$ as the latter estimate is biased due to shot noise, which the former avoids under the assumption that data and randoms are uncorrelated.

\subsection{Implementation}
This method is implemented in \texttt{mockfactory}\footnote{\url{https://github.com/cosmodesi/mockfactory/blob/main/mockfactory/blinding/catalog.py}} which is an MPI-parallel Python package that bundles utility functions to transform a cubic box simulation to a realistic observation mock, in particular, to reproduce the DESI observation. Although the official DESI blinding pipeline \cite{Andrade2024} does not use it, \texttt{mockfactory} also offers an implementation of the BAO and RSD blinding described in \cite{Brieden2020} under the same MPI framework.


\section{Validation with Mocks} 
\label{sec:test}
In this section we use simulations to validate the blinding scheme derived in the previous section. In particular, we first validate the theoretical description of the shot noise contribution, and then the proposed correction to remove this contribution from the blinding scheme. Finally, we emphasize the scales where the blinding scheme is effective and check the distribution of the blinding weights.

\subsection{Mocks} \label{sec:mocks}
To validate the PNG blinding scheme we use a \texttt{FastPM} \cite{Feng2016} simulation generated to validate the DESI PNG analysis\footnote{We used the python version of \texttt{FastPM}: \url{https://github.com/echaussidon/fastpm-python}}. We followed the parametrization of \cite{Ding2022}. This simulation is a box of $5.52$ $h^{-1}$Gpc side length with 6000 meshes per side, \textit{i.e.}, $2.16 \cdot 10^{11}$ particles leading to a dark matter particle mass of about $10^{11}$ M$_\odot$. The initial dark matter field was generated at $z=19$ following the Planck18 cosmology \cite{Planck18} and using the "unitary method", which sets the initial white noise used to generate the initial density field to one, thus reducing the cosmological variance in the simulation \cite{Angulo2016}. Then, the particles evolved up to $z=1.5$ within 40 time steps. In the following, we use the snapshot $z=1.75$.

From this snapshot, the halos are extracted with the popular Friend-of-Friend algorithm \cite{Huchra1982,Press1982}  with a linking length of 0.2 times the mean interparticle distance. In particular, we use a slightly improved version of the MPI code \cite{Feng2017}. Then, to emulate the linear bias of the eBOSS quasars sample observed in~\cite{Laurent2017}, we select all halos with a mass larger than $2.25 \cdot 10^{12}\; M_{\odot}$, which corresponds to at least 178 dark matter particles.

Within this large box, the density of objects is $n = 1.05 \cdot 10^{-3} \; h^{3}\textrm{Mpc}^{-3}$. This full sample will also be denoted as a high-density sample. Two additional subsamples from this full sample are drawn to test the shot noise correction. The first one, of intermediate density, is obtained via a $10 \times$ subsampling, yielding $n = 1.05 \cdot 10^{-4} \; h^{3}\textrm{Mpc}^{-3}$. The second one, of low density, is obtained by a $20 \times$ subsampling, corresponding to $n = 5.24 \cdot 10^{-4} \; h^{3} \textrm{Mpc}^{-3}$. The shot noise is expected to be negligible in the high-density sample, while the low-density sample represents the typical density of the DESI QSO \cite{SV,EDR,Chaussidon2023}, where the shot noise cannot be neglected.

The power spectrum is measured with \texttt{pypower}\footnote{\url{https://github.com/cosmodesi/pypower}} that implements the optimal FFT-based estimator of \cite{Hand2017}.

\subsection{Model vs. mocks without shot noise corrective factor}
\begin{figure}
    \centering
    \includegraphics[scale=.9, center]{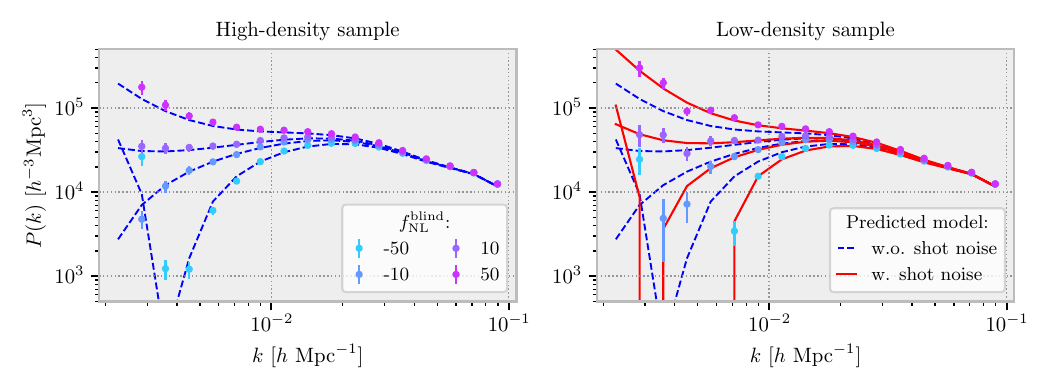}
    \caption{Blinding scheme without the shot noise correction factor, where the shot noise contribution is added to the prediction. Dots are the monopoles of the measured power spectra and the errors are the theoretical prediction for the boxsize at the considered density. Left (\textit{resp.} right) is for the high (\textit{resp.} low) density sample. Blue dashed lines are the monopoles the desired power spectra with the corresponding values of $f_{\mathrm{NL}}^{\mathrm{loc}}$ without shot noise contribution, \textit{i.e.}, \cref{eqn:power_spectrum_with_blinding_weights} in the limit $n \rightarrow \infty$. The red ones are the prediction from including the shot noise contribution \textit{i.e.} \cref{eqn:power_spectrum_with_blinding_weights} with $n=1.05 \cdot 10^{-3} \; h^{3} \textrm{Mpc}^{-3}$ for left and $n=5.24 \cdot 10^{-5} \; h^{3} \textrm{Mpc}^{-3}$ for right. For simplicity, the predictions with the shot noise contribution are not displayed on the left since the blue dashed lines are almost identical.}
    \label{fig:theo_vs_model}
\end{figure}

Let us first apply the blinding scheme without adding the shot noise correction factor and validate the theoretical description given by \cref{eqn:power_spectrum_with_blinding_weights}. The blinded power spectra with different values of $f_{\mathrm{NL}}^{\mathrm{blind}}$ are displayed in \cref{fig:theo_vs_model} for the high-density sample (left) and the low-density sample (right).  

The blue dashed lines are the expected power spectra for the corresponding $f_{\rm NL}^{\rm loc}$ values (with a linear bias adjusted by eye) from \cref{eqn:power_spectrum_with_blinding_weights} in the limit $1/n \rightarrow 0$ \textit{i.e.} without the shot noise contribution. This is what the blinding should reproduce. The red lines correspond to \cref{eqn:power_spectrum_with_blinding_weights} with the shot noise contribution. On the left, the blinding scheme is applied to a high-density sample where the shot noise can be neglected, and we do not display the corresponding red lines since they overlap almost perfectly with the blue lines. The blinding scheme is correctly described in this high-density configuration without the shot noise correction.

For the low-density sample (right), the blinding scheme's impact on the power spectrum is accurately described only when the shot noise correction is included in the model prediction. Once the blinding applied, the blinded data will be fitted with the theory given in \cref{eqn:scale_depedent_bias} without any shot noise correction such that not applying the correction at the blinding level yields a larger PNG signal. Hence, a coherent analysis of data split by their magnitudes or redshifts is made impossible if the shot noise varies across the survey since the blinding is applied only once to the entire sample.

\subsection{Validation of the shot noise correction factor} \label{sec:validation_shotnoise}
\begin{figure}
    \centering
    \includegraphics[scale=0.9, center]{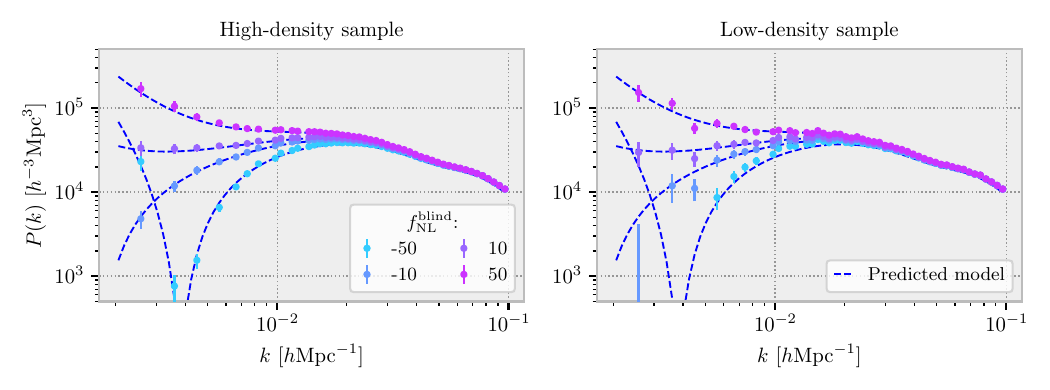}
    \caption{Blinding with the shot noise correction factor for several values of $f_{\rm NL}^{\mathrm{blind}}$ for high-density sample (left) and low-density sample (right). The blue dashed lines are the monopoles of the predicted power spectra with the large scale-dependent bias for the corresponding $f_{\rm NL}^{\mathrm{loc}} = f_{\rm NL}^{\mathrm{blind}}$ given by \cref{eqn:scale_depedent_bias} and the errors are the theoretical prediction for the boxsize at the considered density.. In both cases, blinding with the shot noise correction reproduces the expected shape of the power spectrum.}
    \label{fig:validation_blinding_scheme}
\end{figure}

Let us test the final blinding scheme, which includes the shot noise correction factor. The blinded power spectra are shown in \cref{fig:validation_blinding_scheme} for high and low-density samples. Here, the predicted models are the power spectrum with the large scale-dependent bias given in \cref{eqn:scale_depedent_bias} for $f_{\rm NL}^{\mathrm{loc}} = f_{\rm NL}^{\mathrm{blind}}$ (with a linear bias adjusted by eye). The blinding procedure recovers the desired shape of the power spectrum on large scales. In particular, the shot noise correction, given our choice of $k_p$, works as expected up to the minimal scale of the fitting range \textit{i.e.} $k_{\rm min} \sim 4 \cdot 10^{-3} \; h\textrm{Mpc}^{-1}$. 

One can validate this implementation by fitting the effective PNG signal introduced by the blinding scheme. We use the same parametrization as in the latest eBOSS measurement~\cite{Castorina2019, Mueller2021, Cagliari2023} where the power spectrum is expended on Legendre multipoles
\begin{equation}
    P_{\mathrm{theo}, \ell}(k)=\dfrac{2 \ell+1}{2} \int_{-1}^1 d \mu P_{\mathrm{theo}}(k, \mu) \mathcal{L}_{\ell}(\mu),
\end{equation}
with $P_{\rm theo}(k, \mu)$ a simple model including the Kaiser effect, a damping factor for small scales, the scale-dependent bias, and a shot noise contribution:
\begin{equation} \label{eqn:chap5:pk_theo}
    P_{\mathrm{theo}}(k, \mu)=\dfrac{\left[b + \dfrac{b_{\Phi}}{\alpha(k, z_{\mathrm{eff}})} f_{\mathrm{NL}}^{\mathrm{loc}} + f \mu^2\right]^2}{\left[1+\frac{1}{2}\left(k \mu \Sigma_s\right)^2\right]^{2}} \times P_{\mathrm{lin}}(k, z_{\mathrm{eff}}) + s_{n,0}.
\end{equation}

The fits are performed with \texttt{iminuit} \cite{iminuit}, and posterior samples are drawn with the \texttt{zeus} sampler \cite{Karamanis2021, Karamanis2021a} wrapped into the  \texttt{desilike}\footnote{\url{https://github.com/cosmodesi/desilike}} framework\footnote{A simple example of this model in this framework is given here:\url{https://github.com/echaussidon/desilike/blob/main/nb/png_examples.ipynb}}.
We fit only the monopole and the quadrupole for $k \in [0.004~h\textrm{Mpc}^{-1}, 0.08~h\textrm{Mpc}^{-1}]$ with a binning width of $0.002~h\textrm{Mpc}^{-1}$. The impact of the blinding scheme on the high-order multipoles is shown in \cref{sec:quad-hex}. We fix $b_{\Phi}$ with the universal mass function assumption $b_\phi = 2\delta_c\times(b - 1)$. Finally, the power spectrum covariance is computed analytically in the Gaussian approximation without window effect taking into account the DESI DR1 QSO number density, volume and linear bias.

\begin{figure} 
    \centering
    \includegraphics[scale=0.65, center]{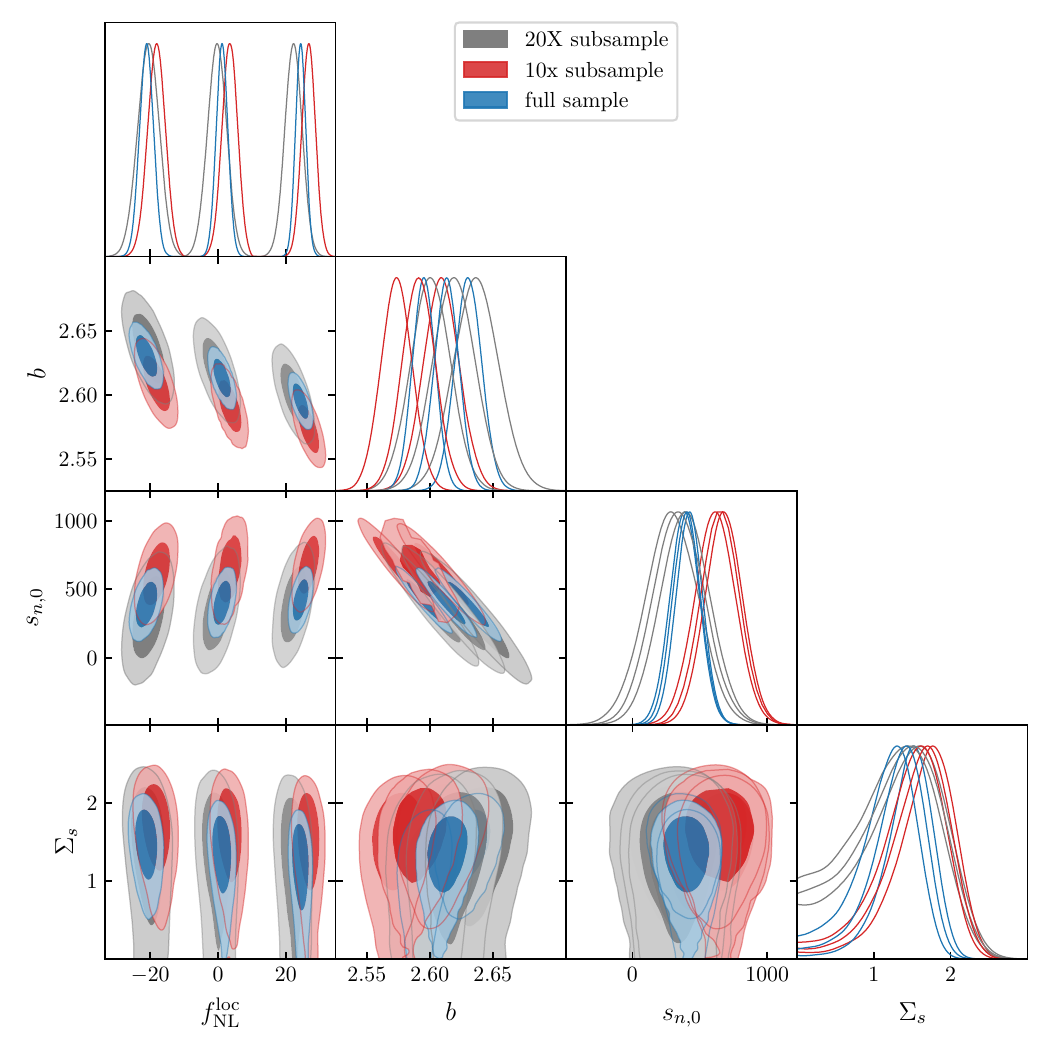}
    \caption{Posteriors obtained for non-blinded ($f_{\rm NL}^{\rm loc} = 0$) and blinded (with $f_{\rm NL}^{\rm loc} \in \{-20, 20\}$) power spectra. Posteriors for the high-density sample are shown in blue, the intermediate-density sample in red, and the low-density sample in grey. As discussed in \cref{sec:lower_bias}, the measured bias changes significantly depending on the $f_{\rm NL}^{\rm blind}$ value.}
    \label{fig:box_with_rsd_blind}
\end{figure}

\begin{figure} 
     \centering
     \begin{subfigure}{\textwidth}
         \centering
         \includegraphics[width=\textwidth]{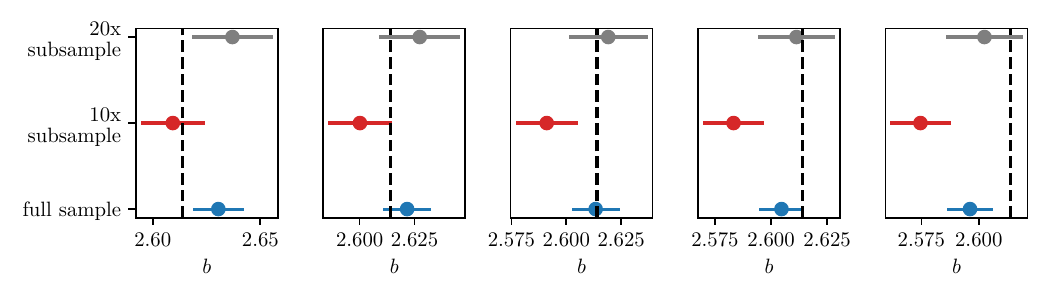}
     \end{subfigure}
     \begin{subfigure}{\textwidth}
         \centering
         \includegraphics[width=\textwidth]{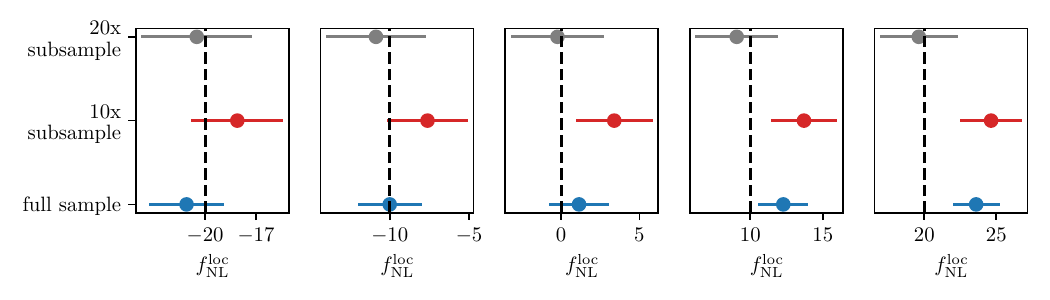}
     \end{subfigure}
        \caption{$b$ and $f_{\rm NL}^{\rm loc}$ best fit values for the non-blinded case ($f_{\rm NL}^{\rm loc} = 0$) and different input values of $f_{\rm NL}^{\rm blind} \in \{-20, -10, 10, 20\}$, with $1\sigma$ posterior errors (see~\cref{fig:box_with_rsd_blind}). Black dashed lines are the expected value of the linear bias and the expected blinded value $f_{\mathrm{NL}}$.}
        \label{fig:fit_dispersion}
\end{figure}

We fit the high-density sample (full sample) and the low-density sample ($20 \times$ subsampling), as well as the intermediate-density sample ($10 \times$ subsampling). The posteriors of $(f_{\rm NL}^{\rm loc}, b, s_{n,0}, \Sigma_s)$ are displayed in \cref{fig:box_with_rsd_blind} and the best-fit values with the errors from the posteriors for $f_{\rm NL}^{\rm loc}$ and $b$ in \cref{fig:fit_dispersion}. 
Note that, as expected, the errors decrease with the density of the sample.

As illustrated in \cref{fig:box_with_rsd_blind}, fitted $f_{\rm NL}^{\rm loc}$ are consistent with input $f_{\rm NL}^{\rm blind}$ value for all density samples. 

\subsection{Blinded range of scales} \label{sec:lower_bias}
\begin{figure}[!h]
    \centering
    \includegraphics[scale=1, center]{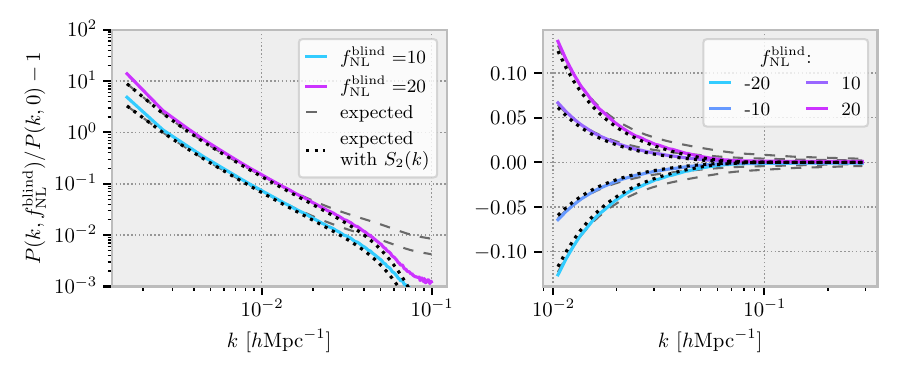}
    \caption{Both figures represent the ratio between the blinded power spectrum monopole and the non-blinded power spectrum monopole for different values of $f_{\rm NL}^{\rm blind}$. The measured power spectra are from the high-density sample to avoid any mismatch due to the shot noise term. Dotted (\textit{resp.} dashed) lines are the model predictions multiplied by the smoothing kernel $S_{2}(k)$ (\textit{resp.} not multiplied by $S_{2}(k)$) from the reconstruction smoothing and mesh assignment. The smoothing kernel damps the PNG signal, such that it goes to zero faster than the theory.} 
    \label{fig:test-model}
\end{figure}

As illustrated in \cref{fig:fit_dispersion}, there is a shift in the measured bias as a function of the $f_{\rm NL}^{\rm blind}$ value. This discrepancy can be understood by examining the ratio between the blind and non-blind power spectrum monopoles displayed in \cref{fig:test-model}. The scale-dependent bias generated by the blinding scheme, given in \cref{eqn:power_spectrum_with_blinding_weights} is altered by $S_{2}(k)$ that erases the small scales. The measured 'high'-$k$ PNG signal is therefore damped by the kernel $S_{2}(k)$, as explained in \cref{sec:model}.

The dashed lines in \cref{fig:test-model} are the theoretical prediction without $S_{2}(k)$ and the black dotted lines are the theoretical prediction ($P(k, f_{\rm NL}^{\rm loc}) / P(k, 0)  - 1$) multiplied by $S_{2}(k)$. Note that this an approximation since $S_{2}(k)$ also has a square contribution\footnote{See the right term in \cref{eqn:power_spectrum_with_blinding_weights}.} leading to the small discrepancy at $k \sim 7 \cdot 10^{-2} \; h\textrm{Mpc}^{-1}$ between the prediction (black dotted lines) and the measured blinding power spectrum (colored lines). Hence, the blinding scheme slightly changes the amplitude of the power spectrum leading to a smaller effective bias.  


The contribution of $S_{2}(k)$ is not included in the fitted model used in \cref{sec:validation_shotnoise}, thereby explaining the different bias values displayed in \cref{fig:box_with_rsd_blind}. This is not considered as an issue, as the galaxy bias is typically marginalized over, and the blinding scheme simulates the correct value of $f_{\rm NL}^{\rm loc}$ within error bars.

Another crucial point is
the impact of the $f_{\rm NL}^{\rm loc}$ blinding is then negligible beyond $k \sim 0.06 \; h\textrm{Mpc}^{-1}$, which should reduce its impact on BAO or RSD measurements, as demonstrated in \cref{sec:impact_bao_rsd}.

\subsection{Distribution of blinding weights}

Since this blinding scheme relies on applying additional weights to the data catalog, the variations in these weights must be small enough compared to those of other typical weights (e.g. to correct for observational systematics) to not provoke accidental unblinding.

The expected standard deviation of the blinding weights, see \cref{eqn:final_weights}, for constant density $n$ is
\begin{equation} \label{eq:dispersion_blinding_weights}
    \begin{aligned}
        \sigma_{w_{\phantom{d}\mathrm{blind}}^{d}}^2 &= \langle (1 + \dfrac{a^{\prime}(k_p, \mu_p, n)}{a(k_p)} w_{\mathrm{blind}} - \langle 1 + \dfrac{a^{\prime}(k_p, \mu_p, n)}{a(k_p)}w_{\mathrm{blind}} \rangle)^2 \rangle \\
       &= \left(\dfrac{a^{\prime}(k_p, \mu_p, n)}{a(k_p)} \right)^2 \langle w_{\mathrm{blind}}(\vb{x}) w_{\mathrm{blind}}(\vb{x}) \rangle  \\
       &=\left(\dfrac{a^{\prime}(k_p, \mu_p, n)}{a(k_p)} \right)^2 \dfrac{1}{(2 \pi)^6} \int \dd \vb{k} \int \dd \vb{q}~e^{i(\vb{k} + \vb{q})\vb{x}} a(k)a(q) \langle \widehat{\delta}^r(\vb{k}) \widehat{\delta}^r(\vb{q}) \rangle  \\
       &= \left(\dfrac{a^{\prime}(k_p, \mu_p, n)}{a(k_p)} \right)^2 \dfrac{1}{2 \pi^2} \int \dd k ~ k^2 a(k)^2 S_2(k)^2\left(b^2P_{\mathrm{lin}}(k) + 1/n \right),
    \end{aligned}
\end{equation}
where in the last equation we have approximated the contribution of $\left(b + (1-S_1(k))f\mu^2\right)$ to $b$, in order to perform the integration. 

Since the simulation box size is of finite size, the integration has to be performed only from $k_{\rm min} = k_{f} = 2 \pi / L$ with $L$ the box size to $k_{\rm max} = k_{\rm N}$ the Nyquist frequency.\footnote{In fact, $k_{\rm max}$ does not really matter since the integrand goes to $0$ really quickly for high $k$.}

\begin{figure}
    \centering
    \includegraphics[scale=1, center]{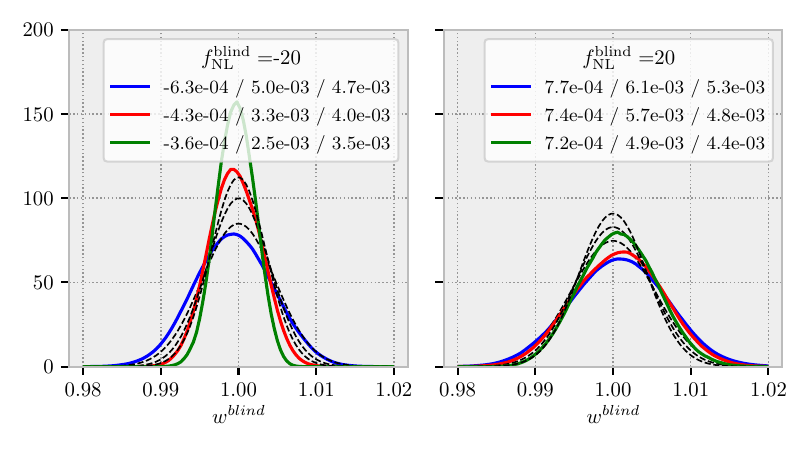}
    \caption{Distribution of the blinding weights in colored lines from the high/intermediate/low-density sample in blue/red/green and the normal distribution with dispersion computed with \cref{eq:dispersion_blinding_weights} in dashed black lines. The shift and standard deviation measured from the mocks, and predicted standard deviation are given in the legend.} 
    \label{fig:weights_dispersion}
\end{figure}

\Cref{fig:weights_dispersion} shows the dispersion of the blinding weights for the corresponding values of $f_{\rm NL}^{\rm loc}$ ($-20$ on the left and $20$ on the right) for the different density samples. Note that the high density sample has a larger dispersion in blinding weights. This is expected since $a^\prime / a$ is bigger for higher density sample as displayed in the lower right panel of \cref{fig:discriminant}.

For each configuration, the shift and standard deviation measured from the mocks and as predicted from \cref{eq:dispersion_blinding_weights} are given in the legend. 
The blinding weights have tiny offset (less than $10^{-3}$) and very small standard deviation (less than $10^{-2}$) such that they can be discreetly mixed with other weights without breaking the blinding procedure. Indeed, for comparison, realistic photometric systematic weights for DESI DR1 data
have a standard deviation of $\sim 5 \cdot 10^{-2}$, see \cite{Ross2024}.  

\section{Test with realistic cutsky mocks} \label{sec:realistic_test}
The ultimate aim of this blinding procedure is to perform systematic and consistency tests on the data without any confirmation bias. Therefore, we have to check whether this method works also in the case of non-zero underlying PNG and if the added fake signal is insensitive to large angular scale imaging systematics and their mitigation. In the following, the blinding scheme is tested with realistic simulations that mimic the DESI DR1 QSO sample.


\subsection{Realistic cutsky mocks}


To build a more realistic simulation, the cubic box $(x, y, z)$ is transformed into a \emph{cutsky} geometry $(\mathrm{R.A.}, \mathrm{Dec.}, z)$ where the density, the redshift distribution, the sky mask, as well as the completeness are matched to the expected DESI QSO DR1 sample \cite{Chaussidon2023}. All of these steps were performed with \texttt{mockfactory}. Note that despite the large size of this box ($5.52 \; h^{-1}\mathrm{Gpc}$), one can only emulate either the North Galatic Cap (NGC) or the South Galactic Cap (SGC) of the DESI Survey. In the following, we decide to emulate the SGC part of the survey.

As mentioned above, the DESI QSO density corresponds to that of the low-density sample. Hence, 16 different subsamples can be extracted from the full sample. Although they are not entirely independent, we use the average on these 16 subsamples to reduce shot noise.

In the following, instead of using the analytical Gaussian approximation for the covariance matrix, we estimate it from 1000 EZMock fast simulations that match the DESI QSO DR1 SGC sample in the same manner as the cutsky \texttt{FastPM}. These EZMocks generated for DESI are very similar to the ones described in \cite{Zhao2021} (see \cite{Zhang2023} for an assessment of the covariance matrix from EZMocks). 
 
Additionally, to account for the effect of the survey mask in the measured power spectra, all the fits are performed by convolving the theoretical power spectrum by the corresponding window function as explained in \cite{Beutler2021}. The window functions are also computed with \texttt{pypower}. We do not account for the negligible global integral constraint for this test.

\subsection{Blinding with non-zero PNG simulations} \label{sec:non_zero_sim}
First, we test the blinding procedure in the presence of a non-zero PNG signal. 
Similarly to the $f_{\rm NL}^{\rm loc} = 0$ \texttt{FastPM} simulation described in \cref{sec:mocks}, we have also produced one simulation with $f_{\rm NL}^{\rm loc} = 25$, from which 16 subsamples matching the DESI QSO DR1 SGC are extracted following the procedure described above. This simulation was produced with the same initial conditions, \textit{i.e.}, using the same initial density field, that is rescaled for the PNG case, to generate the initial particle distribution, see \cite{Angulo2022}. The fits are performed on the mean of these 16 subsamples without rescaling the mock-based covariance matrix. 

\begin{figure}
    \centering
     \begin{subfigure}{0.47\textwidth}
         \centering
         \includegraphics[scale=0.9, center]{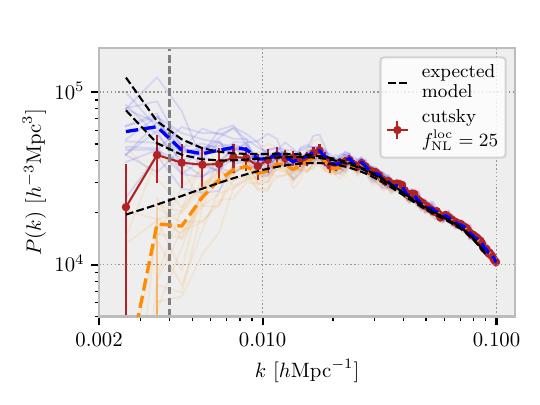}
         \caption{}
         \label{fig:cutsky_non_zero_power}
     \end{subfigure}
     \hfill
     \begin{subfigure}{0.47\textwidth}
         \centering
         \includegraphics[scale=1, center]{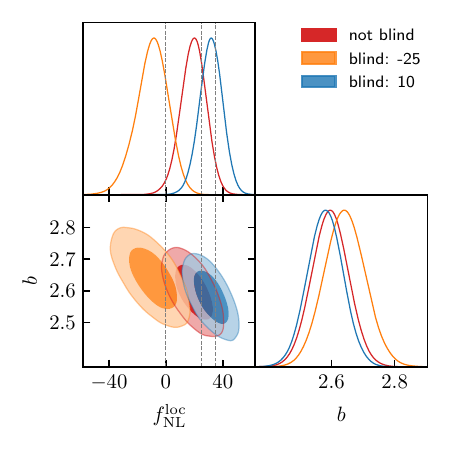}
         \caption{}
         \label{fig:cutsky_non_zero_mcmc}
     \end{subfigure}
    \caption{(a) The brick line is the mean (over the 16 subsamples) measured power spectrum monopole without blinding for the DR1 SGC footprint, based on \texttt{FastPM} simulations with $f_{\mathrm{NL}}^{\mathrm{loc}}=25$, with error bars from 1000 EZMocks. The monopoles from each subsample are display in the background. Orange and blue lines show the power spectrum monopoles blinded with $f_{\mathrm{NL}}^{\mathrm{loc}} \in \{-25, 10\}$ respectively. The dashed vertical line is about the minimum scale used in the fit. (b) Posteriors for the three power spectrum monopoles in (a). The posterior of the \texttt{FastPM} (non-blinded) power spectrum is slightly shifted to a lower $f_{\mathrm{NL}}^{\mathrm{loc}}$ value than input, which propagates to the blinded cases.}
    \label{fig:cutsky_non_zero}
\end{figure}

The DR1 SGC power spectra for the non-blinded case and blinding values of $f_{\rm NL}^{\rm blind} \in \{-25, 10\}$ and the corresponding posteriors are displayed in \cref{fig:cutsky_non_zero}. The best-fit values are $19_{-9}^{+9}$ (non-blinded), $31_{-8}^{+9}$ (blinded + 10) and  $-9_{-10}^{+12}$ (blinded - 25) where the errors are the $1\sigma$ credible intervals. Note that the errors are smaller than expected for the real DESI analysis since the value of $b_\phi$ is computed with $p=1$, see \cite{Slosar2008}. We mostly recover the expected behavior for the blinded power spectra down to the scale of interest ($k_{\rm min} = 4 \cdot 10^{-3} \; h~\textrm{Mpc}^{-1}$). Note that there is some small difference between the non-blinded power spectrum and the model, which is propagated to the blinded cases.

This difference in the non-blinded case is from the fact that halos in our \texttt{FastPM} simulations are extracted with a FoF halo finder which is known to produce halos that are not correctly described by the universal relation used for $b_{\Phi}$, see \citep{Desjacques2009, Biagetti2017} for an extensive study. One explanation could be that FoF algorithm creates non-spherical halos, while the universal relation for $b_{\Phi}$ is derived from only spherical halos. This discrepancy will not impact our study since our simulation with $f_{\rm NL}^{\rm loc} = 25$ has just a lower apparent amount of PNG signal.

\subsection{Blinding with a mispecified survey selection function (remaining systematics)} \label{sec:ultimate_test}
\begin{figure}
    \centering
     \begin{subfigure}{0.47\textwidth}
         \centering
         \includegraphics[scale=1, center]{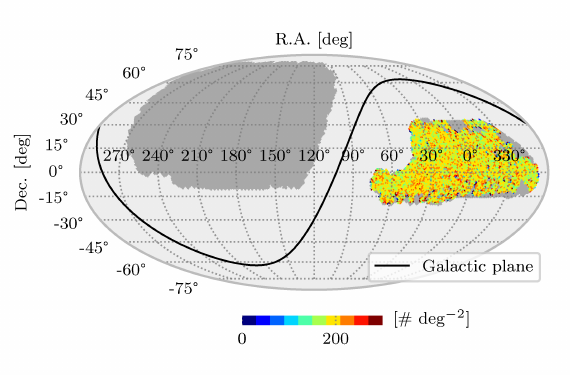}
         \caption{}
         \label{fig:cutsky_non_zero_power}
     \end{subfigure}
     \hfill
     \begin{subfigure}{0.4\textwidth}
         \centering
         \includegraphics[scale=1, center]{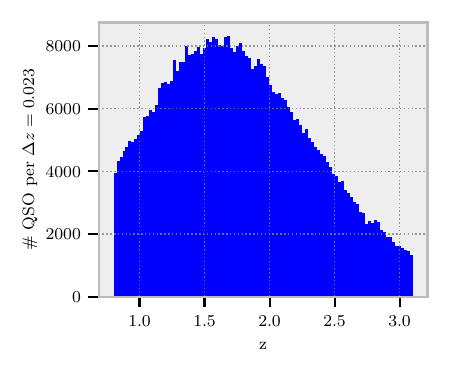}
         \caption{}
         \label{fig:cutsky_non_zero_mcmc}
     \end{subfigure}
    \caption{(a) Angular density distribution of one subsample of the \texttt{FastPM} simulation contaminated by realistic imaging systematics. The dark grey region is the expected final DESI footprint. (b) Redshift distribution of the same subsample.}
\label{fig:survey_selection_function}
\end{figure}

The ultimate test is to ensure that the blinding scheme is not altered by the presence of imaging systematics and by the mitigation applied for the correction. For this purpose, each subsample is contaminated by realistic imaging systematics. The footprint and the redshift distribution of one of this subsample are shown in \cref{fig:survey_selection_function}. Based on the density fluctuation in the sky observed in the DESI DR1 sample, we start with a subsample with a higher density than observed and we remove real objects in each pixel on the sky to match the density fluctuation at \texttt{HEALPix} \cite{Gorski2005} level with $n_{side}=256$. We then apply the blinding scheme, and finally the imaging systematics are corrected either with the inverse of the contamination or with \texttt{regressis}\footnote{\url{https://github.com/echaussidon/regressis}} \citep{Chaussidon2022}. The measured value of $f_{\mathrm{NL}}^{\mathrm{loc}}$ is expected to recover the input $f_{\mathrm{NL}}^{\mathrm{blind}}$.

\begin{figure}
    \centering
    \includegraphics[width=0.9\textwidth, center]{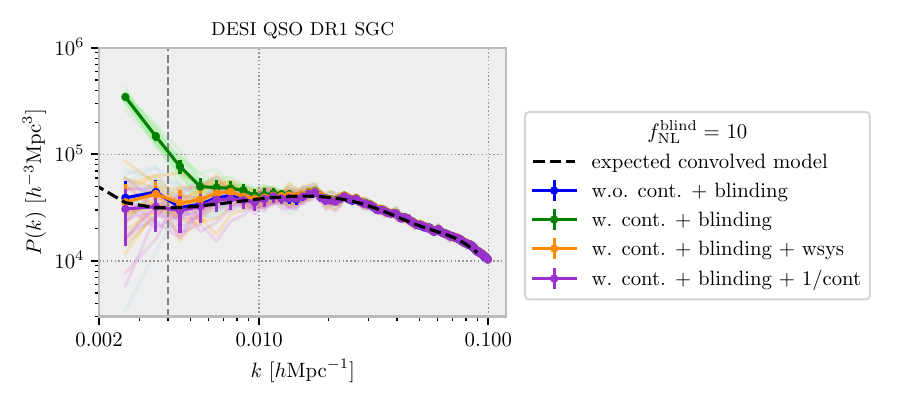}
    \caption{Mean of 16 subsamples mimicking the DESI QSO DR1 SGC sample. The errors are from the standard deviation of 1000 EZMocks representing the SGC. The dashed vertical line is at the minimum scale used in the fit. The blue line is the power spectrum monopole measured in simulations without imaging systematics and with blinding. Green is for contaminated simulations with blinding. Orange (\textit{resp.} violet) is for contaminated simulations with blinded and correction with \texttt{regressis} (\textit{resp.} inverse of the contamination).}
    \label{fig:power_spectrum_ultimate}
\end{figure}

The means of the 16 subsamples of the power spectra for the different cases are shown in \cref{fig:power_spectrum_ultimate}. 
Note the importance of the imaging systematic mitigation in extracting the PNG signal.  

\begin{figure}
     \centering
     \includegraphics[scale=1, center]{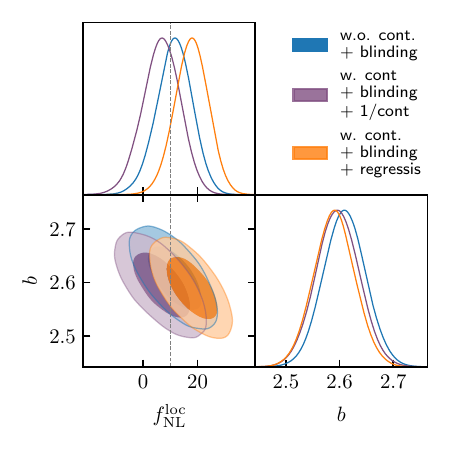}
     \caption{Blue contours correspond to the blinding applied on the uncontaminated sample. Violet (\textit{resp.} orange) contours correspond to blinding applied on the contaminated sample and with correction weights given by the inverse of the contamination (\textit{resp.} with \texttt{regressis}). We almost recover the input blinding value: $f_{\mathrm{NL}}^{\rm blind} = 10$ in the three cases. Here, we fit the power spectrum from SGC cutsky with the full DR1 covariance matrix containing both NGC and SGC, leading to expected uncertainties for the DR1 analyses.}
     \label{fig:mcmc_ultimate}
\end{figure}

$f_{\mathrm{NL}}^{\mathrm{loc}}$ posteriors, for the mean of the 16 subsamples, are displayed in \cref{fig:mcmc_ultimate}. Here, we use the covariance matrix from the DR1 footprint instead of the one from the SGC footprint to reflect DR1 uncertainty. 
Note that in the fits we used $p=1.0$ since this is the value used during the blinding step. However, in the case of the QSO, one certainly wants to use $p=1.6$ instead of $p=1.0$ \cite{Slosar2008} during the fit, such that the expected errors with DESI DR1 will be a little bit larger than these ones. We recover the input blinding value $f_{\mathrm{NL}}^{blind} = 10$ well within the DR1 uncertainty both if no contamination is applied and including the imaging systematics, with the exact inverse correction. In the latter case, we obtain a slightly lower value of $f_{\mathrm{NL}}^{\mathrm{loc}}$. The correction with \texttt{regressis} is slightly farther off ($1.4 \sigma$), but no fine-tuning of \texttt{regressis} was performed at this stage such that the imaging weights computed here are not optimal.

\newpage
\section{Conclusion} \label{sec:conclusion}
We have developed a method to blind the PNG signal that manifests as a large scale-dependent bias in the power spectrum. The proposed blinding scheme provides a set of weights to be applied to the data or the randoms such that the density field is altered at large scales to mimic a chosen scale-dependent bias. In particular, we have shown how shot noise propagates into our blinding scheme and how we can correct for it. This correction is significant and ensures a consistent fake PNG signal within a sample with varying density. We have also verified (in \cref{sec:impact_bao_rsd}) 
that the addition of this blinding does neither alter the measurement of the BAO and RSD parameters at the statistical accuracy of DR1 measurements nor alter the blinding setup for BAO and RSD.

As a result, this blinding was introduced into DESI's clustering analysis pipeline, and since forecasts predict a sensitivity on $f_{\mathrm{NL}}^{\mathrm{loc}}$ of  $10$ to $15$ at $68\%$ of confidence level with the DR1 data, we drew a single blinded random value for all the DESI tracers $f_{\rm NL}^{\rm blind} \in [-15, 15]$~\cite{Andrade2024}.

Although it was not the purpose here, the blinding scheme proposed could be tested and adapted to work for any other promising summary statistics as the CMB lensing \cite{Krolewski2023}, skew spectra \cite{Dizgah2020,Dai2020} or bispectrum \cite{Jeong2009}. Note that this blinding scheme will only alter the scale-dependent bias and is not able to blind the PNG signal coming from the matter density field. This is particularly relevant since, for instance, CMB lensing is not sensitive to the same systematics as the galaxy power spectrum and can help to provide an unbiased measurement. Finally, we emphasize also that this method can be easily adapted to any search of signal that changes the bias, for instance, the detection of relativistic contributions to galaxy clustering \cite{Beutler2020}.

\section*{Data Availability}
The blinding scheme is publicly available here: \url{https://github.com/cosmodesi/mockfactory}, as well as all the other tools mentioned in this article: \url{https://github.com/cosmodesi/}.
All the material needed to reproduce all the figures of this publication is available here: \url{https://doi.org/10.5281/zenodo.11270652}.

\acknowledgments
This material is based upon work supported by the U.S. Department of Energy (DOE), Office of Science, Office of High-Energy Physics, under Contract No. DE–AC02–05CH11231, and by the National Energy Research Scientific Computing Center, a DOE Office of Science User Facility under the same contract. Additional support for DESI was provided by the U.S. National Science Foundation (NSF), Division of Astronomical Sciences under Contract No. AST-0950945 to the NSF’s National Optical-Infrared Astronomy Research Laboratory; the Science and Technology Facilities Council of the United Kingdom; the Gordon and Betty Moore Foundation; the Heising-Simons Foundation; the French Alternative Energies and Atomic Energy Commission (CEA); the National Council of Science and Technology of Mexico (CONACYT); the Ministry of Science and Innovation of Spain (MICINN), and by the DESI Member Institutions: \url{https://www.desi.lbl.gov/collaborating-institutions}. Any opinions, findings, and conclusions or recommendations expressed in this material are those of the author(s) and do not necessarily reflect the views of the U. S. National Science Foundation, the U. S. Department of Energy, or any of the listed funding agencies.

The authors are honored to be permitted to conduct scientific research on Iolkam Du’ag (Kitt Peak), a mountain with particular significance to the Tohono O’odham Nation.

\bibliographystyle{biblio_style}
\bibliography{bibli}

\appendix
\section{Weights applied on data or randoms?} \label{sec:weights_data_randoms}
Blinding weights can be applied either on the data: $1 + w_{\rm blind}(\vb{x})$ or on the randoms: $1 - w_{\rm blind}(\vb{x})$. For the latter configuration, the FKP field reads
\begin{equation} \label{eqn:FKP_field_randoms}
\begin{aligned}
    F(x) &= W(x) \left(1 + \delta_g(x)\right) - W(x) \left( 1 - w_{\rm blind}(x) \right) \\ 
         &= W(x) \delta_g(x) + W(x) w_{\rm blind}(x).
\end{aligned}
\end{equation}

\begin{figure}[!h]
    \centering
    \includegraphics[scale=1.0, center]{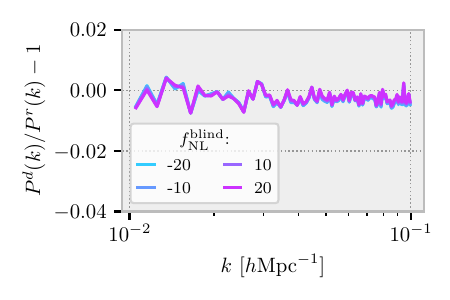}
    \caption{Ratio between the power spectrum measured from the blinded low-density sample, with blinding weights applied to the data and to the randoms, for different $f_{\rm NL}^{\rm blind}$ values.}
    \label{fig:randoms_vs_data_weights}
\end{figure}

To recover the expected FKP field in \cref{eqn:FKP_field}, we have neglected the contribution of $\delta_g(x) w_{\mathrm{blind}}(x)$ that does not appear in \cref{eqn:FKP_field_randoms}. The difference between the two cases is negligible, as shown in the \cref{fig:randoms_vs_data_weights} for the low-density subsample, where the difference is expected to be the highest.

To keep the definition of randoms as sampling the survey selection function\footnote{Then randoms are the same for blind and non-blind data.}, with as little correlation as possible, we decided to apply the blinding weights to the data.

\section{Impact of blinding scheme on the quadrupole and hexadecapole}\label{sec:quad-hex}
As described in \cref{eqn:power_spectrum_with_blinding_weights}, the blinding scheme is expected to modify the quadrupole ($\ell=2$). However, the hexadecapole ($\ell=4$) in the linear regime is $P_{4}(k) = \frac{8}{35} f^{2} P_{\mathrm{lin}}(k)$ and does not depend on the bias such that the blinding procedure should not modify it.

\begin{figure}
    \centering
    \includegraphics[scale=.9, center]{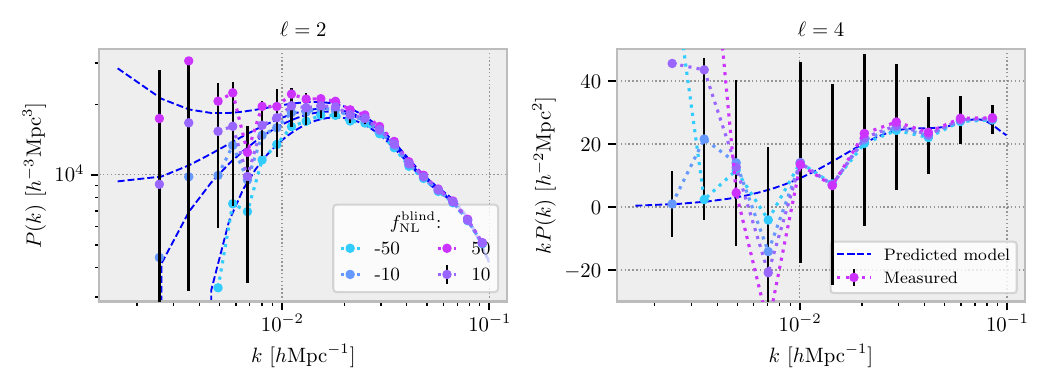}
    \caption{Quadrupole (left) and hexadecapole (right) for the high-density sample with the blinding scheme applied with the shot noise correction for several values of $f_{\rm NL}^{\rm blind}$. The corresponding monopole is displayed on the left panel of \cref{fig:validation_blinding_scheme}. The blue dashed lines are the predicted power spectra with the scale dependent bias and adjusted linear bias. For simplicity, the gaussian errors (black lines) are only displayed for $f_{\rm NL}^{\rm blind}=10$.}
    \label{fig:hexadecapole}
\end{figure}

The quadrupole and hexadecapole for multiple values of $f_{\rm NL}^{\rm blind}$ and for the high-density sample are displayed in \cref{fig:hexadecapole}. We recover the expected behavior for both quadrupole and hexadecapole. Note that the statistical power at large scales for the hexadecapole is very weak and our measurements match well enough the expected theoretical prediction without PNG signal.

\section{Impact on the BAO / RSD measurement} \label{sec:impact_bao_rsd}
As explained in \cref{sec:lower_bias}, the window function related to the reconstruction prevents small scales from being impacted by the PNG blinding scheme. Hence, we quantify here the impact of this blinding scheme on the BAO/RSD measurement and how it can be incorporated with the BAO and the RSD blinding given in \cite{Brieden2020}. In this section, we are using the DESI blinding pipeline described in \cite{Andrade2024}\footnote{\url{https://github.com/desihub/LSS/blob/main/scripts/main/apply_blinding_main_fromfile_fcomp.py}}.

For this test, we used the 25 mock catalogs from the \texttt{AbacasSummit} N-body simulations \cite{Maksimova2021, Garrison2021} that are transformed into light-cone to mimic the DESI DR1 Luminous Red Galaxy sample (LRG) \cite{Zhou2023} and use the full redshift range ($0.4 < z < 1.1$). The associated covariance with this sample is produced by \texttt{TheCov}\footnote{\url{https://github.com/cosmodesi/thecov}} \cite{Wadekar2020,Kobayashi2023,Alves2024}. We use the monopole, quadrupole and hexadecapole for $0.02 < k [h\textrm{Mpc}^{-1}] < 0.2$ with a binning of $\mathrm{d}k = 0.005~h\textrm{Mpc}^{-1}$.  

We perform a standard full shape analysis using the 1-loop power spectrum computed by \texttt{velocileptors}\footnote{\url{https://github.com/sfschen/velocileptors}} \citep{Chen2020, Chen2021}. As usual, we parameterize the Alcock-Paczynski (AP) effect \cite{Alcock1979} with $q_{\|}, q_{\perp}$ such that the power spectrum in the true cosmological coordinates is related to the power spectrum in observable coordinates by
\begin{equation}
P^{\mathrm{obs}}\left(\vb{k}_{\mathrm{obs}}\right)=q_{\perp}^{-2} q_{\|}^{-1} P(\vb{k}), \quad k_{\|, \perp}^{\mathrm{obs}} = q_{\|, \perp} k_{\|, \perp},
\end{equation}
with $q_{\|}$ and $q_{\perp}$ the shifts along and perpendicular to the line-of-sight, respectively.
We denote $X^{\rm fid}$ the value of $X$ in the fiducial cosmology used to transform redshift into distances. See, for instance, \cite{Maus2024} for a complete description. 

In addition to the AP parameters $(q_{\|}, q_{\perp})$, we also fit the growth rate $f$ via $df = f/f^{\rm fid}$ and the ShapeFit parameter $m$. The parameter $df$ is extracted thanks to the RSD terms. The ShapeFit parameter $m$ was introduced by \cite{Brieden2021a} in order to collect additional information thanks to the shape of the power spectrum. It modifies the linear power spectrum as
\begin{equation}
P_{\rm lin}(k)=P_{\rm lin}^{\rm fid}(k) \exp \left(\frac{m}{a} \tanh \left[a \ln \left(\frac{k}{k_{piv}}\right)\right]\right),
\end{equation}
where we use $a=0.6$ \cite{Brieden2021a} and $k_{piv} = \pi / r_d$ is the pivot scale. $P_{\rm lin}^{\rm fid}$ is the linear power spectrum of the fiducial theory that is fixed during all the fit.

We test several different blinding configurations based on two BAO/RSD blinding configurations\footnote{See \cite{Andrade2024} for a description of the BAO/RSD blinding values.}: blind-1 $(w_0, w_a = -1.2, 0.75)$ and blind-2 $(w_0, w_a = -0.8, -0.75)$, for which we test a PNG blinding value of $f_{\rm NL}^{\rm blind} \in [-15, 0, 15]$. For comparison, we also test, for the same PNG blinding value, a configuration without the BAO/RSD blinding.

\begin{figure}
    \centering
    \includegraphics[scale=1, center]{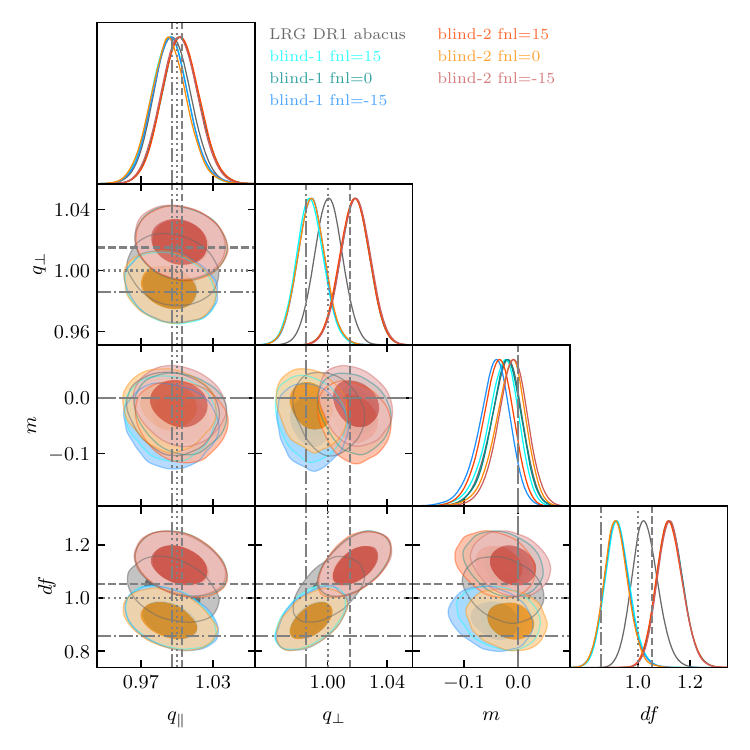}
    \caption{Posteriors from the mean of 25 LRG \texttt{Abacus} mocks for the different blinding configurations. Gray is for unbind results, while blue color is for blind-1 and red is for blind-2. Although the BAO/RSD measurements on blinding mocks are imperfect, the PNG blinding scheme does not impact $q_{\|}, q_{\perp}, df$. $m$ is slightly impacted, see \cref{fig:ratio_to_unblind}.}
    \label{fig:final_fit_bao_rsd}
\end{figure}

The posteriors are drawn from the mean power spectrum of the 25 mocks to reduce the cosmic variance, but we use the covariance for the DR1 sample. They are obtained with \texttt{emcee} \cite{Foreman-Mackey2013} and displayed in \cref{fig:final_fit_bao_rsd}. 
Note that we have a larger value of $df$ than expected due to wrong velocities for this version of simulation. This was corrected in the latter versions, however it will not impact the discussion here. Adding the PNG blinding appears to have no impact on the BAO/RSD parameters and slightly impacts the ShapeFit parameter $m$, as already noticed with simulations with PNG signal \cite{Brieden2021} since the PNG blinding changes the large-scale slope of the power spectrum. 

\begin{figure}
    \centering
    \includegraphics[scale=0.9, center]{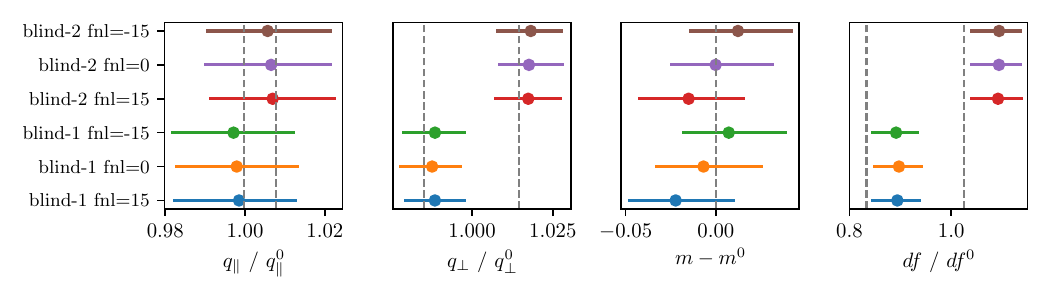}
    \caption{Ratios or difference between the parameters measured with the different blinded configurations and with the non-blinded case where the parameters are computed as the mean in the MCMC chains, see \cref{fig:final_fit_bao_rsd}, and the errors are the $1\sigma$ credible intervals from the chains in the blinded cases. The gray lines are the ratio between the expected value from the blinding and the ones measured in the non-blinded case.}
    \label{fig:ratio_to_unblind}
\end{figure}

To quantify the deviation, the ratios between the parameters measured with the different blinded configurations and with the non-blinded case are plotted in \cref{fig:ratio_to_unblind}. The parameters are computed as the mean in the MCMC chains, while the errors are the $1\sigma$ credible interval from the chains in the blinded cases. The gray lines are the ratios between the expected value from the blinding and those measured in the non-blinded case. 

We note that the PNG blinding scheme, as expected, does not impact the scales where the BAO and RSD parameters $(q_{\|}, q_{\perp}, df)$ are measured and, therefore, can be included without biasing the further analysis of these measurements. The discrepancy on $m$ is expected, as explained in \cite{Brieden2021}. It is not a major issue however, as shifts remain below the statistical uncertainty, and the blinding procedure primarily aims at hiding $(q_{\|}, q_{\perp}, df)$ and $f_{\rm NL}^{\rm loc}$.  

\section{Author Affiliations}
\label{sec:affiliations}

\noindent \hangindent=.5cm $^{1}${Lawrence Berkeley National Laboratory, 1 Cyclotron Road, Berkeley, CA 94720, USA}

\noindent \hangindent=.5cm $^{2}${IRFU, CEA, Universit\'{e} Paris-Saclay, F-91191 Gif-sur-Yvette, France}

\noindent \hangindent=.5cm $^{3}${Physics Dept., Boston University, 590 Commonwealth Avenue, Boston, MA 02215, USA}

\noindent \hangindent=.5cm $^{4}${Department of Physics \& Astronomy, University College London, Gower Street, London, WC1E 6BT, UK}

\noindent \hangindent=.5cm $^{5}${Institute for Computational Cosmology, Department of Physics, Durham University, South Road, Durham DH1 3LE, UK}

\noindent \hangindent=.5cm $^{6}${Instituto de F\'{\i}sica, Universidad Nacional Aut\'{o}noma de M\'{e}xico,  Cd. de M\'{e}xico  C.P. 04510,  M\'{e}xico}

\noindent \hangindent=.5cm $^{7}${Kavli Institute for Particle Astrophysics and Cosmology, Stanford University, Menlo Park, CA 94305, USA}

\noindent \hangindent=.5cm $^{8}${SLAC National Accelerator Laboratory, Menlo Park, CA 94305, USA}

\noindent \hangindent=.5cm $^{9}${Institut d'Estudis Espacials de Catalunya (IEEC), 08034 Barcelona, Spain}

\noindent \hangindent=.5cm $^{10}${Institute of Cosmology and Gravitation, University of Portsmouth, Dennis Sciama Building, Portsmouth, PO1 3FX, UK}

\noindent \hangindent=.5cm $^{11}${Institute of Space Sciences, ICE-CSIC, Campus UAB, Carrer de Can Magrans s/n, 08913 Bellaterra, Barcelona, Spain}

\noindent \hangindent=.5cm $^{12}${School of Mathematics and Physics, University of Queensland, 4072, Australia}

\noindent \hangindent=.5cm $^{13}${Sorbonne Universit\'{e}, CNRS/IN2P3, Laboratoire de Physique Nucl\'{e}aire et de Hautes Energies (LPNHE), FR-75005 Paris, France}

\noindent \hangindent=.5cm $^{14}${Departament de F\'{i}sica, Serra H\'{u}nter, Universitat Aut\`{o}noma de Barcelona, 08193 Bellaterra (Barcelona), Spain}

\noindent \hangindent=.5cm $^{15}${Institut de F\'{i}sica d’Altes Energies (IFAE), The Barcelona Institute of Science and Technology, Campus UAB, 08193 Bellaterra Barcelona, Spain}

\noindent \hangindent=.5cm $^{16}${NSF NOIRLab, 950 N. Cherry Ave., Tucson, AZ 85719, USA}

\noindent \hangindent=.5cm $^{17}${Instituci\'{o} Catalana de Recerca i Estudis Avan\c{c}ats, Passeig de Llu\'{\i}s Companys, 23, 08010 Barcelona, Spain}

\noindent \hangindent=.5cm $^{18}${Departamento de F\'{i}sica, Universidad de Guanajuato - DCI, C.P. 37150, Leon, Guanajuato, M\'{e}xico}

\noindent \hangindent=.5cm $^{19}${Instituto Avanzado de Cosmolog\'{\i}a A.~C., San Marcos 11 - Atenas 202. Magdalena Contreras, 10720. Ciudad de M\'{e}xico, M\'{e}xico}

\noindent \hangindent=.5cm $^{20}${Department of Physics and Astronomy, University of Waterloo, 200 University Ave W, Waterloo, ON N2L 3G1, Canada}

\noindent \hangindent=.5cm $^{21}${Perimeter Institute for Theoretical Physics, 31 Caroline St. North, Waterloo, ON N2L 2Y5, Canada}

\noindent \hangindent=.5cm $^{22}${Waterloo Centre for Astrophysics, University of Waterloo, 200 University Ave W, Waterloo, ON N2L 3G1, Canada}

\noindent \hangindent=.5cm $^{23}${Instituto de Astrof\'{i}sica de Andaluc\'{i}a (CSIC), Glorieta de la Astronom\'{i}a, s/n, E-18008 Granada, Spain}

\noindent \hangindent=.5cm $^{24}${Center for Cosmology and AstroParticle Physics, The Ohio State University, 191 West Woodruff Avenue, Columbus, OH 43210, USA}

\noindent \hangindent=.5cm $^{25}${Department of Astronomy, The Ohio State University, 4055 McPherson Laboratory, 140 W 18th Avenue, Columbus, OH 43210, USA}

\noindent \hangindent=.5cm $^{26}${The Ohio State University, Columbus, 43210 OH, USA}

\noindent \hangindent=.5cm $^{27}${Department of Physics and Astronomy, Sejong University, Seoul, 143-747, Korea}

\noindent \hangindent=.5cm $^{28}${CIEMAT, Avenida Complutense 40, E-28040 Madrid, Spain}

\noindent \hangindent=.5cm $^{29}${Department of Physics, University of Michigan, Ann Arbor, MI 48109, USA}

\noindent \hangindent=.5cm $^{30}${University of Michigan, Ann Arbor, MI 48109, USA}

\noindent \hangindent=.5cm $^{31}${Department of Physics \& Astronomy, Ohio University, Athens, OH 45701, USA}

\noindent \hangindent=.5cm $^{32}${National Astronomical Observatories, Chinese Academy of Sciences, A20 Datun Rd., Chaoyang District, Beijing, 100012, P.R. China}

\end{document}